\documentclass[12pt]{article}

\usepackage{axodraw}
\usepackage{color}

\usepackage{epsfig}

\def\beq{\begin{equation}}
\def\eeq{\end{equation}}
\def\bea{\begin{eqnarray}}
\def\eea{\end{eqnarray}}

\def\gev{\, {\rm GeV}}
\def\mev{\, {\rm MeV}}

\newcommand{\gsim}{\lower.7ex\hbox{$\;\stackrel{\textstyle>}{\sim}\;$}}
\newcommand{\lsim}{\lower.7ex\hbox{$\;\stackrel{\textstyle<}{\sim}\;$}}

\def\mwhat{\hat m_W}
\def\mzhat{\hat m_Z}
\def\glhat{\hat \Gamma_{l^+l^-}}
\def\s2eff{\hat s^2_{\rm eff}}
\def\ahat{\hat\alpha}

\def\gfhat{\hat G_F}
\def\oexpt{\hat {\cal O}^{\rm expt}}
\def\doexpt{\Delta \hat {\cal O}^{\rm expt}}
\def\otheory{{\cal O}^{\rm th}}
\def\predict#1{{\rm Prediction~of}~#1 }
\def\tp#1{({#1})^{\rm th}}

\def\SetColor#1{}

\begin{document}

\setlength{\baselineskip}{0.25in}


\begin{titlepage}
\noindent
\begin{flushright}
MCTP-04-74 \\
December 2004 \\
\end{flushright}
\vspace{1cm}

\begin{center}
  \begin{Large}
    \begin{bf}
{\large TASI Lecture Notes} \\ 
Introduction to Precision Electroweak Analysis

    \end{bf}
  \end{Large}
\end{center}
\vspace{0.2cm}
\begin{center}
\begin{large}
James D. Wells \\
\end{large}
  \vspace{0.3cm}
  \begin{it}
Michigan Center for Theoretical Physics (MCTP) \\
~~University of Michigan, Ann Arbor, MI 48109-1120, USA \\
\vspace{0.1cm}
\end{it}

\end{center}

\begin{abstract}

I give a basic introduction to precision electroweak analysis, beginning
with calculation at tree-level which most simply illustrates the
procedure.  I then work out the formalism for one-loop corrections to
the vector boson self energies (oblique corrections).  This is a
tractable subset of the complete electroweak program.  Not only does
this exercise provide an analytically accessible demonstration of the
theory involved in electroweak precision analyses, it also teaches
students a useful technique to analyze a large class of theories
beyond the Standard Model.

\end{abstract}

\vspace{1cm}

\begin{center}
{\it Presented at ``Theoretical Advanced Study Institute in Elementary
Particle Physics (TASI 2004): Physics in $D\geq 4$,'' Boulder, Colorado,
6 Jun -- 2 Jul, 2004.}
\end{center}


\end{titlepage}

\section{Symmetries, dynamics and observables}

Voltaire said that natural philosophy is about calculating
and measuring -- almost everything else is chimera.  I suppose our three
lectures on precision electroweak analysis will be with Voltaire's
blessing, as it involves much calculating and measuring.
We focus mostly on the calculating activity in these lectures,
and rely on the measuring results produced by our experimental colleagues.

Our discussion starts with observables. 
We have fancy names for observables, such as ``total hadronic cross-section'',
and ``leptonic partial width'' and ``effective weak mixing angle''.
However, observables quantify very tangible events
that happen in nature, albeit we humans forced the action:  
$A$ lit up when $B$ hit it, etc.  That is a rather crude way of
thinking about it, but in the end we must remember that
measurement is about stuff slapping or pulling or yanking other stuff.
Observables are just measurements with the blood wiped off.

Definining an observable in modern elementary particle physics
involves sophistication and a fair amount of theoretical knowledge,
but we will not get into a chicken and egg discussion.  Let us 
suppose we have a nice collection of well-defined observables ${\cal O}_i$,
presented within the context of a theory, and we wish to determine
if it all makes sense.  In other words, we wish to determine if our
theory can explain the observables.
Finding a theory that matches observables is not hard at all.  Give
me any set of $n$ observables $\{ {\cal O}_i\}$ and I can give
you this theory:  For every ${\cal O}_i$ we posit the reason
${\cal R}_i$, which simply states that ${\cal O}_i$ is true.
This is a sort of ``intelligent design'' theory that cannot be
ruled out as untrue but is clearly unsatisfactory and not useful
to modern scientists.

Within particle physics today, we posit that symmetries (relations) and dynamics
(strengths) produce nature, and observables are mere
manifestations of these qualities.  Observables have relations among themselves
dictated by the symmetries and dynamics.
Nature automatically gets all the relations right,
and there is no intermediary needed by nature to calculate results.
However, we humans need an intermediate step to understand the relations
of observables.  We (usually) need a lagrangian field theory to
tie the symmetries and dynamics to the observables.  The drawing on the next page
visualizes this connection.

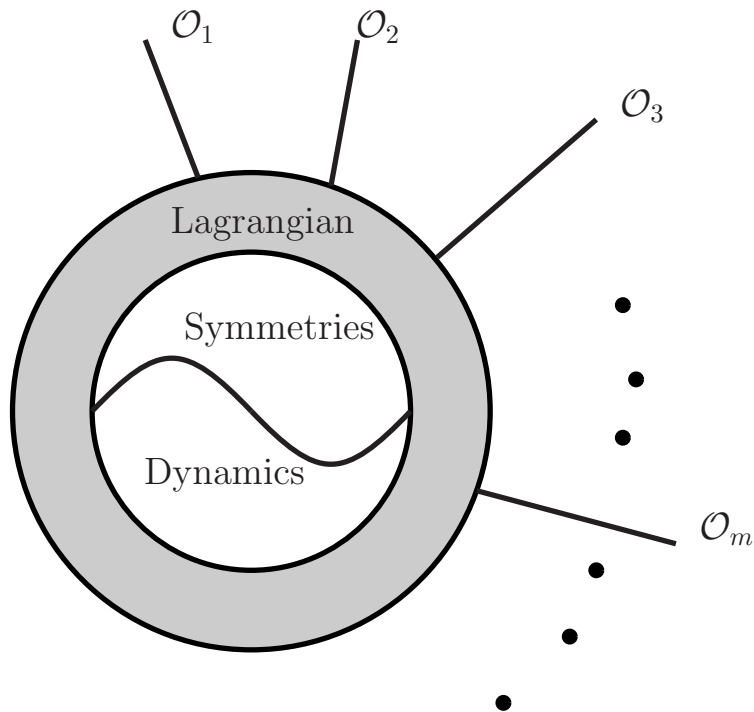
\begin{center}
\begin{picture}(350,300)(0,0)
\SetWidth{2.0}
\GOval(130,130)(90,90)(0){0.8}
\GOval(130,130)(60,60)(0){1}
\Line(110,218)(90,270)
\Line(160,215)(170,270)
\Line(200,188)(260,240)
\Line(215,100)(290,80)
\Photon(70,130)(190,130){20}{1}
\Text(100,270)[lb]{\large ${\cal O}_1$}
\Text(170,270)[lb]{\large ${\cal O}_2$}
\Text(270,240)[lb]{\large ${\cal O}_3$}
\Text(300,80)[lb]{\large ${\cal O}_m$}
\GOval(270,170)(2,2)(0){0}
\GOval(275,142)(2,2)(0){0}
\GOval(270,120)(2,2)(0){0}
\GOval(260,70)(2,2)(0){0}
\GOval(250,45)(2,2)(0){0}
\GOval(225,20)(2,2)(0){0}
\Text(105,155)[lb]{\large Symmetries}
\Text(90,100)[lb]{\large Dynamics}
\Text(100,195)[lb]{\large Lagrangian}
\end{picture}
\end{center}

The symmetries and dynamics that we posit in these lectures are the
ones of the Standard Model. At the core of the SM is its gauge groups
$SU(3)\times SU(2)_L\times U(1)_Y$ and their corresponding
gauge coupling strengths, $g_3$, $g_2$ and $g'$.  

At first we perform a tree-level analysis of the SM.  This will
be used to show how the lagrangian is a mere catalyst to relating
observables in terms of observables.  We will be able to show that
the predicted relations at tree level are not satisfied by the data.
Our next topic will be to discuss one loop corrections to the tree-level
result.  We will focus on the interesting subset of vector boson self-energies,
which is illustrative of the general method and useful in beyond-the-SM
analyses.  We then will compute an example of the one-loop corrections
due to fermion loops in the vector boson-self energies.  We will show that
when we write observables in terms of observables, there are no infinities.
They cancel out automatically -- we do not have to subtract them, they just
are not there in physical calculations.  Finally, we describe the utility
of the techniques that we develop in these few short lectures.  It is
not all just for pedagogy.  There is some direct use to what we say for
analysing some theories beyond the Standard Model.

\section{Tree-level analysis of SM precision electroweak observables}

We will use the notation that observables are written with
a hat on top of them. For example, we will denote 
the measured $Z$ boson mass observable as $\mzhat$. 
The observables that we are primarily interested
in are $\ahat$ (from Thomson limit of $\gamma^*\to e^+e^-$ scattering),
$\gfhat$ (from muon decay), $\mzhat$ ($Z$ boson mass), $\mwhat$ ($W$ boson
mass),  $\glhat$ (leptonic
partial width of the $Z$ boson), and
$\s2eff$ (effective $\sin^2\theta_W$).  The value of $\s2eff$ is defined
to be the all-orders rewriting of $\hat A_{LR}(= A_e)$ as
\beq
\hat A_{LR}\equiv \frac{(1/2-\s2eff)^2-\hat s^4_{\rm eff}}
                 {(1/2-\s2eff)^2+\hat s^4_{\rm eff}}.
\eeq
The measured values of all these observables\cite{PDG,LEPEWWG} are
\bea
\ahat & = & 1/137.0359895(61) \\
\gfhat & = & 1.16639(1)\times 10^{-5}\gev^{-2} \\
\mzhat & = & 91.1875\pm 0.0021 \gev  \\
\mwhat & = & 80.426\pm 0.034\gev  \\
\s2eff & = & 0.23150 \pm 0.00016  \\
\glhat & = & 83.984\pm 0.086\mev
\eea

At tree level we need only three lagrangian parameters to compute
the six observables listed above. The three parameters are
$g$ ($SU(2)$ gauge coupling), $g'$ ($U(1)_Y$ gauge coupling) and
$v$ (Higgs vacuum expectation value).  In anticipation of the convenience
we will wish upon our one-loop
discussion later, we cash in these three parameters for an equivalent
set $e$, $s (=\sin\theta)$, and $v$, where $g=e/s$ and $g'=e/c$.

We can now compute all the observables in terms of these three lagrangian
parameters:
\bea
\ahat & = & \frac{e^2}{4\pi} \\
\gfhat & = & \frac{1}{\sqrt{2}v^2} \\
\mzhat^2 & = & \frac{e^2v^2}{4s^2c^2} \\
\mwhat^2 & = & \frac{e^2 v^2}{4s^2} \\
\s2eff & = & s^2 \\
\glhat & = & \frac{v}{96\pi}\frac{e^3}{s^3c^3}
\left[  \left(-\frac{1}{2}+2s^2\right)^2+\frac{1}{4}\right]
\eea

The equations above have measurements on the LHS and theory computations in terms
of lagrangian parameters on the RHS.  Staring at these equations enables us
to scoff at questions like, {\it Does the SM predict the correct $W$ mass?}
Well, yes it does, if that's the only observable we care about. We have an
infinite number of ways (choices of $e$, $s$, and $v$) to reproduce the $W$
mass.  The real question that a theory must answer is, {\it Can we reproduce
all experimental results with suitable choices of our input parameters?}
This is a serious question requiring analysis.

The standard way to test the ability of a theory to reproduce data is via
the $\chi^2$ analysis.  We have a set of observables $\oexpt_i$
with uncetainties $\doexpt_i$. The theory makes predictions $\otheory_i$
for the observables that depend on the lagrangian parameters.  We find
the best possible choices of the lagrangian parameters that fit the data
by minimizing the $\chi^2$ function
\beq
\chi^2(e,s,v) = \sum_i \frac{ (\oexpt_i-\otheory_i(e,s,v))^2}{(\doexpt_i)^2}
\eeq
where $i$ sums over the observables $\mwhat$, $\s2eff$, etc.
A good discussion of how to interpret the statistics of the $\chi^2$
distribution can be found in the PDG\cite{PDG}.  

In our list of six observables, three of them are measured extraordinarily
well: $\ahat$, $\gfhat$ and $\mzhat$.  We can get a feel for how well 
the tree-level SM predictions match data by fixing the lagrangian parameters
$e$, $s$ and $v$ in terms of these three observables, and then writing
the remaining observables in terms of the 
$\ahat$, $\gfhat$ and $\mzhat$.

With simple algebra we find that
\bea
e^2 & = & {4\pi \ahat} \\
v^2 & = & \gfhat^{-1}/\sqrt{2} \\
s^2c^2 & = & \frac{\pi \ahat}{\sqrt{2}\gfhat \mzhat^2}\label{s2c2} 
\eea
The last equation is equivalent to
\beq
s^2  =  \frac{1}{2}-\frac{1}{2}\sqrt{1-4\hat x}~~{\rm where} ~~
\hat x = \frac{\pi \ahat}{\sqrt{2}\gfhat \mzhat^2}
\eeq

We can now write the observables $\mwhat$, $\s2eff$ and $\glhat$
in terms of $\ahat$, $\mzhat$ and $\gfhat$:
\beq
\mwhat^2 =  \pi\sqrt{2}\gfhat^{-1}\ahat\left( 1
-\sqrt{1-\frac{4\pi\ahat}{\sqrt{2}\gfhat \mzhat^2}}\right)^{-1} 
\eeq
\beq
\s2eff =  \frac{1}{2}-\frac{1}{2}
\sqrt{1-\frac{4\pi\ahat}{\sqrt{2}\gfhat \mzhat^2}} 
\eeq
\beq
\glhat  =  \frac{\sqrt{2}\gfhat \mzhat^3}{12\pi}
\left\{ \left( \frac{1}{2}-\sqrt{1-\frac{4\pi\ahat}{\sqrt{2}\gfhat \mzhat^2}}
\right)^2+\frac{1}{4}\right\}
\eeq

If we plug in the very precisely known experimental values for $\ahat$,
$\mzhat$ and $\gfhat$, we find predictions for $\mwhat$, $\s2eff$ and $\glhat$:
\bea
\predict{\mwhat} & = & 80.939\pm 0.003\gev \\
\predict{\s2eff} & = & 0.21215\pm 0.00003 \\
\predict{\glhat} & = & 84.843\pm 0.012\mev
\eea
The predictions of $\mwhat$, $\s2eff$ and $\glhat$ in this particular
tree-level procedure are approximately $15\sigma$, $120\sigma$ and $10\sigma$ off
from their experimentally measured values.  Statistically speaking,
these are unacceptably large deviations of the theory from the experiment.
We therefore conclude that the theory is not compatible with experiment.

However, we have only worked up to tree-level in the perturbative expansion
of the theory.  We must go to higher-order in the coupling constants to 
truly test the viability of the SM when confronting all the experimental
data.  This analysis has been applied to the Standard Model, and one finds
that it is compatible
with the precision electroweak data {\em provided} the Higgs
boson mass is between about $114\gev$ (direct bound) and $219\gev$ (95\% C.L.
upper bound)\cite{LEPEWWG}.  

In the following sections we will consider how one-loop
self-energies slightly alter the relationship between 
the lagrangian parameters and
measured observables.  In other words, the relationships among 
observables are slightly
different than what we found above doing a tree-level analysis,
and the theory predictions come closer to the experimental
measurements.

\section{One-loop self-energy corrections}

In this lecture we focus on the class of corrections that arise solely
from the self-energy corrections of the $\gamma$, $W^\pm$, and $Z$
vector bosons.  Restricting our analysis to this class of corrections
enables us to do something complete and meaningful in the short time
we have together. A full-scale renormalization of the SM with all corrections
explicitly calculated is a significantly more time-consuming project
without significantly enhancing the conceptual learning.  Furthermore,
many of the most interesting ideas of physics beyond the SM require
only analysing self-energy corrections to the vector bosons.  For
example, additional exotic states that do not couple directly to the
SM fermions but have charges under the SM gauge symmetries qualify to
be analysed in this manner.

Even in beyond-the-SM theories which have exotic states that do
interact with the external fermions involved in precision electroweak
analysis, it is most common that the non-oblique corrections have a
small effect compared to the oblique corrections. This is generally
true in supersymmetry, with the notable exception of the $Z\to b\bar
b$ coupling that participates in $R_b$, $A_b$ and $A_{FB}^b$,
which can (but generically does not) have large vertex corrections due
to superpartners.  One main reason for the dominance of oblique
corrections over non-oblique corrections is that any charged object
couples to the vector bosons, whereas usually only one or two
particles in a theory couple to a specific fermion species.  In other
words, the summing over all contributors in self-energies wins out
over the one or two diagrams that couple to an individual final state
fermion.

To begin our analysis we stipulate that
the lagrangian is the SM lagrangian and 
the couplings that affect the precision electroweak
observables $\s2eff$, $\mwhat$, and $\glhat$ are $\{ e,s^2,v^2\}$.  
The relevant Feynman rules
for our analysis are 

\begin{center}
\begin{picture}(260,200)(-10,0)
\SetWidth{1.0}
\Photon(20,50)(70,50){3}{4}
\Photon(20,100)(70,100){3}{4}
\Photon(20,150)(70,150){3}{4}
\ArrowLine(70,150)(100,165)
\ArrowLine(70,100)(100,115)
\ArrowLine(70,50)(100,65)
\ArrowLine(100,135)(70,150)
\ArrowLine(100,85)(70,100)
\ArrowLine(100,35)(70,50)
\Text(150,45)[lb]{\large $\frac{ie}{s\sqrt{2}}\gamma_\mu P_L$}
\Text(150,95)[lb]{\large $\frac{ie}{sc}\gamma_\mu
                  \left[ (T_f^3-Q_fs^2)P_L-Q_f s^2P_R\right]$}
\Text(150,145)[lb]{\large $ieQ_f\gamma_\mu$}
\Text(110,160)[lb]{\large $f$}
\Text(110,130)[lb]{\large $\bar f$}
\Text(110,110)[lb]{\large $f$}
\Text(110,80)[lb]{\large $\bar f$}
\Text(110,60)[lb]{\large $\mu^-$}
\Text(110,30)[lb]{\large $\bar \nu$}
\Text(-3,140)[lb]{\large $A_\mu$}
\Text(-3,90)[lb]{\large $Z_\mu$}
\Text(-3,40)[lb]{\large $W^-_\mu$}
\end{picture}
\end{center}

By convention the one-loop corrections to the vector boson self-energies 
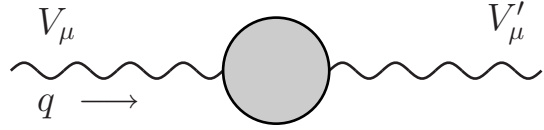
\begin{center}
\begin{picture}(210,70)(-10,0)
\SetWidth{1.0}
\Photon(0,30)(80,30){3}{4}
\Photon(120,30)(200,30){3}{4}
\GOval(100,30)(20,20)(0){0.8}
\Text(10,40)[lb]{\large $V_\mu$}
\Text(180,40)[lb]{\large $V'_\mu$}
\Text(10,12)[lb]{\large $q~\longrightarrow$}
\end{picture}
\end{center}
is of the form
\beq
i[\Pi_{VV'}(q^2)g^{\mu\nu}-\Delta_{VV'}(q^2)q^\mu q^\nu].
\eeq

Only the $\Pi_{VV'}$ piece of the self-energies matters for our 
analysis since the $q^\mu$ part
of the second term is dotted into a light-fermion current and is 
zero by the Dirac equation, since the
corresponding fermion masses is well-approximated to be zero:
\beq
q^\mu J_\mu^{\rm light\, fermion}\rightarrow 
\bar f \gamma^\mu q_\mu f\rightarrow
\bar f m f\rightarrow 0.
\eeq
The way the self-energies are defined, they add to the vector boson masses by convention:
\bea
m^2_V\to m^2_V+\Pi_{VV}(q^2=m^2_V) 
 \eea

Because the photon is massless we know that $\Pi_{\gamma\gamma}(0)=0$
and $\Pi_{\gamma Z}(0)=0$, and so we do not have to compute them.
There is one subtlety to keep in mind.  $\Pi_{\gamma Z}(0)$ is not
zero when the $W^\pm$ bosons is included in the loop.  The procedure
that we outline below gets slightly more complicated when we take that
into account, and the details of that procedure can be found, for example, 
in several studies\cite{Hollik:1988ii,Kennedy:1992tj}.  However, this is special 
to the $W^\pm$ bosons (gauge degree
of freedom partners of the $W^3$).  In new physics scenarios (e.g.,
supersymmetry) there are no additional one-loop contributions to
$\Pi_{\gamma Z}(0)$, and it is usually appropriate in analyses
of beyond-the-SM contributions to precision EW observables to ignore
it.

\subsection{Theoretical predictions for observables at one loop}

The computation of the $Z$ and $W$ masses is straightforward.  
The resulting theoretical
prediction of $m_Z$ and $m_W$ in terms of the lagrangian parameters 
and the one-loop
self-energy corrections is
\beq
\tp{\mzhat}=\frac{e^2 v^2}{4s^2c^2}+\Pi_{ZZ}(m^2_Z)
\eeq
\beq
\tp{\mwhat}=\frac{e^2v^2}{4s^2}+\Pi_{WW}(m^2_W)
\eeq

We next compute the theory prediction for $\alpha$.  It sounds odd to 
use the words ``theory prediction of $\alpha$'' 
since we often are sloppy in our wording (or thinking) and view $\alpha$ as
just a coupling.  In reality, it is an observable defined in the 
Thomson limit of Compton
scattering and probes the Coulomb potential at $q^2\to 0$:
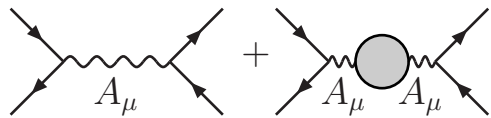
\begin{center}
\begin{picture}(200,55)(120,0)
\SetWidth{1.0}
\ArrowLine(150,30)(130,10)
\ArrowLine(130,50)(150,30)
\ArrowLine(190,30)(210,50)
\ArrowLine(210,10)(190,30)
\ArrowLine(230,50)(250,30)
\ArrowLine(250,30)(230,10)
\ArrowLine(290,30)(310,50)
\ArrowLine(310,10)(290,30)
\Photon(150,30)(190,30){2}{4}
\Photon(250,30)(260,30){2}{2}
\Photon(280,30)(290,30){2}{2}
\GOval(270,30)(10,10)(0){0.8}
\Text(218,28)[lb]{\large $+$}
\Text(162,10)[lb]{\large $A_\mu$}
\Text(248,10)[lb]{\large $A_\mu$}
\Text(277,10)[lb]{\large $A_\mu$}
\end{picture}
\end{center}
which is proportional to 
\beq
-i\left. \frac{4\pi\hat \alpha}{q^2}\right|_{q^2\to 0}
=\frac{-ie^2}{q^2}\left[ 1 +\frac{\Pi_{\gamma\gamma}(q^2)}{q^2}\right]_{q^2\to 0}
\eeq
If we define 
\beq
\Pi'_{\gamma\gamma}(0)\equiv \lim_{q^2\to 0}\frac{\Pi_{\gamma\gamma}(q^2)}{q^2}
\eeq
then we can write the theory prediction for $\alpha$ as
\beq
\tp{\ahat}=\frac{e^2}{4\pi}\left( 1 + \Pi'_{\gamma\gamma}(0)\right)
\eeq

The muon decay observable $\gfhat$ is computed from the lifetime of the muon

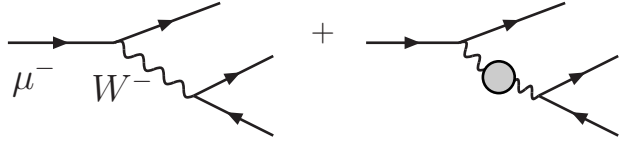
\begin{center}
\begin{picture}(243,80)(107,0)
\SetWidth{1.0}
\ArrowLine(110,55)(150,55)
\ArrowLine(150,55)(190,70)
\ArrowLine(180,35)(210,50)
\ArrowLine(210,20)(180,35)
\ArrowLine(245,55)(280,55)
\ArrowLine(280,55)(320,70)
\ArrowLine(310,35)(340,50)
\ArrowLine(340,20)(310,35)
\Photon(150,55)(180,35){2}{4}
\Photon(280,55)(290,45){2}{2}
\Photon(300,40)(310,35){2}{2}
\GOval(295,42)(6,6)(0){0.8}
\Text(112,35)[lb]{\large $\mu^-$}
\Text(142,33)[lb]{\large $W^-$}
\Text(225,55)[lb]{$+$}
\end{picture}
\end{center}
which is proportional to $\hat G_F/\sqrt{2}$.
This amplitude is then used to compute the muon lifetime
\beq
\tau_\mu^{-1} = \frac{\gfhat^2m^5_\mu}{192\pi^3}K(\alpha,m_e,m_\mu,m_W)
\eeq
where the function $K$ is mainly a kinematics function and 
can be obtained from the electroweak
chapter in the PDG\cite{PDG}.
The theory prediction for $\gfhat$ is
\bea
\frac{\tp{\gfhat}}{\sqrt{2}} & = & \frac{g^2}{8m^2_W}\left[ 1 +
i\Pi_{WW}(q^2)\left( \frac{-i}{q^2-m_W^2}\right)
\right]_{q\to 0} \nonumber \\
& = &
\frac{1}{2v^2}\left[ 1-
\frac{\Pi_{WW}(0)}{m^2_W}\right].
\eea
 
The observable associated with $\s2eff$ is a little trickier than the
other ones.  For one, there are many different types of $\s2eff$
observables, depending on the final state fermion.  We will define
$\s2eff$ to be the observable associated with the left-right asymmetry
of $Z$ decays to leptons.  We assume universality of the leptons.  The
left-right asymmetry is defined to be the $Z$-pole production
cross-section asymmetry of leptons produced from left polarized
electron-positron collisions versus those produced from right
polarized collisions,
\beq
A_{LR}^l = \frac{\sigma_L-\sigma_R}{\sigma_L+\sigma_R} \equiv \frac{c_L^2-c_R^2}{c^2_L+c^2_R}
\eeq
where at tree-level the $c_L$ and $c_R$ couplings are defined by

\begin{center}
\begin{picture}(260,90)(-10,0)
\SetWidth{1.0}
\Photon(20,50)(70,50){3}{4}
\ArrowLine(70,50)(100,65)
\ArrowLine(100,35)(70,50)
\Text(150,45)[lb]{\large $i\gamma_\mu (c_LP_L+c_RP_R)$}
\Text(110,60)[lb]{\large $f$}
\Text(110,30)[lb]{\large $\bar f$}
\Text(-3,40)[lb]{\large $Z_\mu$}
\end{picture}
\end{center}
and
\beq
c_L=\frac{e}{sc}(T^3-Qs^2)~~~{\rm and}~~~c_R=-\frac{-eQs^2}{sc}
\eeq

The definition of $\s2eff$ is chosen such that observable $\hat A_{LR}^l$ 
is written in terms of $\s2eff$ using the tree-level expression above
with $s^2\to \s2eff$. This is an unambiguous definition since
the charges $Q$ and $T^3$ do not get renormalized.  This definition
of $\s2eff$ will become clearer below as we compute it at one loop.

At this point, we need to compute the one-loop shifts 
in $c_L$ and $c_R$.
We can neglect all $\Pi_{ZZ}$ contributions since they will only 
affect the overall factor
of $c_L$ and $c_R$ which cancels.   On the other hand, the 
$Z-A$ mixing self-energy does contribute to the $c_{L}$ and $c_R$ couplings:

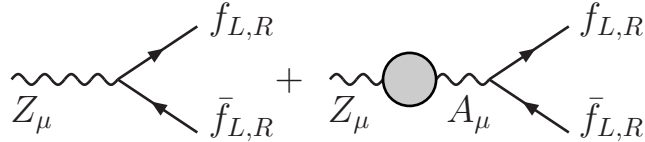
\begin{center}
\begin{picture}(230,85)(100,0)
\SetWidth{1}
\Photon(110,50)(150,50){2}{4}
\Photon(230,50)(250,50){2}{2}
\Photon(270,50)(290,50){2}{2}
\GOval(260,50)(10,10)(0){0.8}
\ArrowLine(150,50)(180,70)
\ArrowLine(180,30)(150,50)
\ArrowLine(290,50)(320,70)
\ArrowLine(320,30)(290,50)
\Text(110,30)[lb]{\large $Z_\mu$}
\Text(185,65)[lb]{\large $f_{L,R}$}
\Text(185,27)[lb]{\large $\bar f_{L,R}$}
\Text(230,30)[lb]{\large $Z_\mu$}
\Text(275,30)[lb]{\large $A_\mu$}
\Text(325,65)[lb]{\large $f_{L,R}$}
\Text(325,27)[lb]{\large $\bar f_{L,R}$}
\Text(210,45)[lb]{\large $+$}
\end{picture}
\end{center}

where
\bea
c_L & = & \frac{e}{sc}(T^3-Qs^2) + i \Pi_{\gamma Z}(m^2_Z) 
      \left( \frac{-i}{m^2_Z}\right) (eQ) \\
 & = & \frac{e}{sc}\left[ T^3 -Q\left(s^2-sc\frac{\Pi_{\gamma Z}(m^2_Z)}{m^2_Z}\right)\right]
 \eea
 \bea
 c_R & = & \frac{-eQs^2}{sc}+i\Pi_{\gamma Z}(m_Z^2)\left( \frac{-i}{m^2_Z}\right) (eQ) \\
 & = & -\frac{eQ}{sc}\left[ s^2-sc \frac{\Pi_{\gamma Z}(m^2_Z)}{m^2_Z}\right].
 \eea

The above $c_L$ and $c_R$ expressions are exactly the same as the 
tree-level expressions
except $s^2\to s^2-sc\Pi_{\gamma Z}(m^2_Z)/m^2_Z$ in the numerator.  
Thus, at the $Z$-pole
\beq
\tp{\s2eff} =s^2-sc\frac{\Pi_{\gamma Z}(m_Z^2)}{m_Z^2} 
\eeq
\beq
{\rm where}~\hat A_{LR} =\frac{(1/2-\s2eff)^2-(\s2eff)^2}
{(1/2-\s2eff)^2+(\s2eff)^2}.
\eeq

Now we compute $\glhat$ from
\begin{center}
\begin{picture}(330,80)(40,0)
\SetWidth{1}
%
\Photon(50,40)(90,40){2}{4}
\Text(50,20)[lb]{\large $Z_\mu$}
\ArrowLine(90,40)(120,60)
\ArrowLine(120,20)(90,40)
\Text(124,35)[lb]{\large $+$}
\Photon(150,40)(170,40){2}{2}
\Photon(190,40)(210,40){2}{2}
\GOval(180,40)(10,10)(0){0.8}
\ArrowLine(210,40)(240,60)
\ArrowLine(240,20)(210,40)
\Text(150,20)[lb]{\large $Z_\mu$}
\Text(190,20)[lb]{\large $A_\mu$}
\Photon(270,40)(290,40){2}{2}
\Photon(310,40)(330,40){2}{2}
\GOval(300,40)(10,10)(0){0.8}
\ArrowLine(330,40)(360,60)
\ArrowLine(360,20)(330,40)
\Text(270,20)[lb]{\large $Z_\mu$}
\Text(310,20)[lb]{\large $Z_\mu$}
\Text(244,35)[lb]{\large $+$}
\end{picture}
\end{center}
The theoretical prediction for this observable in terms of independent 
lagrangian parameters
and one-loop self-energies is
\beq
\tp{\glhat} = \frac{Z_Z}{48\pi}\frac{e^2}{s^2c^2} \mzhat \left[ \left( -\frac{1}{2}+2\tp{\s2eff}\right)^2 +\frac{1}{4}
\right]
\eeq
Recall that $\Pi_{\gamma Z}$ had the effect of just putting 
$s^2\to \tp{\s2eff}$ into the numerator
of the $c_L$ and $c_R$ expressions.  The $\mzhat$ comes as a kinematical 
phase space mass of
the $Z$ decay.

Since we are computing a partial width and not a ratio of couplings, 
the $\Pi_{ZZ}$
contribution must now be taken into account.  The parameter  
$Z_Z$ in the $\tp{\glhat}$ expression
results from this contribution.  It is a wavefunction residue piece.  
To compute this contribution we first
must recognize that the $\Pi_{ZZ}(q^2)$ self-energy when resummed affects the 
$Z$ boson propagator in a simple way
\beq
{\rm Resummed~Propagator}~~
\longrightarrow ~~ 
P^{\mu\nu}_Z(q^2) = \frac{-ig^{\mu\nu}}{q^2-m^2_Z-\Pi_{ZZ}(q^2)}.
\eeq
But,
\bea
\Pi_{ZZ}(q^2) = \Pi_{ZZ}(m^2_{\rm phys})+\Pi'_{ZZ}(m^2_{\rm phys})(q^2-m_{\rm phys}^2)+\cdots
\eea
where $m_{\rm phys}^2$ is really just the physical $Z$ mass, $\mzhat$
(I am writing
$m_{\rm phys}$ here just for emphasis).  The mass of the $Z$ is defined 
to be the position of
the real part of the pole of the propagator.  From that definition and 
the expansion given above,
we find that in the neighborhood of $q^2=m^2_{\rm phys}$
\bea
q^2-m_Z^2-\Pi_{ZZ}(q^2) & = &
q^2-m_Z^2-\Pi_{ZZ}(m_{\rm phys}^2) - \Pi'_{ZZ}(m^2_{\rm phys})
(q^2-m_{\rm phys}^2) +\cdots \nonumber \\
& = & (q^2-m^2_{\rm phys})(1-\Pi'_{ZZ}(m^2_{\rm phys}))+\cdots
\eea
Therefore, in the neighborhood of $q^2=m^2_{\rm phys}$ the $Z$ 
propagator can be written
as
\beq
\frac{-ig^{\mu\nu}}{(q^2-m^2_{\rm phys})(1-\Pi'_{ZZ}(m^2_{\rm phys})}=
\frac{-iZ_Zg^{\mu\nu}}{(q^2-m^2_{\rm phys})}
\eeq
where
\beq
Z_Z=1+\Pi'_{ZZ}(\mzhat )+\, {\rm higher~order~terms}
\eeq

A standard approximation for $\Pi'_{ZZ}(m^2_Z)$ is
\beq
\Pi'_{ZZ}(m^2_Z) =\frac{\Pi_{ZZ}(m^2_Z)-\Pi_{ZZ}(0)}{m^2_Z}
\eeq
This is a good approximation for many scenarios, and we will employ it
hereafter just so we can match up with results published by others.
However, I would like to emphasize that there is no reason why one
needs to use this approximation, especially since there are now many
good numerical and analytic tools to evaluate the one-loop
self-energies.  Sometimes we will also utilize the variable
$\delta_Z$ which is defined as $Z_Z=1+\delta_Z$, where
\beq
\delta_Z = \Pi'_{ZZ}(m^2_Z) \simeq \frac{\Pi_{ZZ}(m^2_Z)-\Pi_{ZZ}(0)}{m^2_Z}
=\frac{\Pi_{ZZ}(m^2_Z)}{m^2_Z}-\frac{\Pi_{ZZ}(0)}{m^2_Z}
\eeq

\subsection{Observables in terms of observables}

At this point we have written all of our observables in terms of
lagrangian parameters and $\Pi$ functions (one-loop corrections).  We
now wish to do some analytical inversions of these expressions and
compute observables in terms of other observables, similar to what we
did in the tree-level analysis at the beginning of these lectures.
There is no need for us to do this in principle.  We are perfectly set
now to compute the one-loop self-energies in our favorite theory and
then try to fit the lagrangian parameters in a total $\chi^2$
analysis.  Indeed, a full renormalization of the SM or any other
theory of equivalent complexity is virtually impossible to
analytically invert in order to write observables in terms of observables.
However, we can do it here, by virtue of the relatively noncomplex
nature of the one loop self-energies.  Furthermore, as emphasized at
the beginning, I wish to do this for pedagogical reasons, to show that 
one role of theories is to be
able to express observables in terms of other observables. 
This knowledge may give one a different perspective about the infinities
that supposedly afflict our theories.

Before we do those calculations, we need to say a few more things 
about the $\ahat$ observable.
It is an unusual observable among our list, because it is 
obviously incalculable.  Recall from before
that we found
\beq
e^2=\frac{4\pi\ahat}{1+\Pi'_{\gamma\gamma}(0)}
\eeq
The problem is with $\Pi'_{\gamma\gamma}(0)$, 
which requires us to know the result of the photon
self energy as $q^2\to 0$:

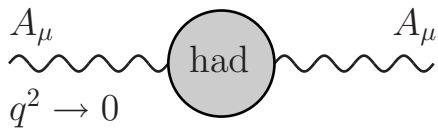
\begin{center}
\begin{picture}(140,50)(-5,-5)
\SetWidth{1.0}
\Photon(0,30)(60,30){3}{4}
\Photon(100,30)(160,30){3}{4}
\GOval(80,30)(20,20)(0){0.8}
\Text(0,6)[lb]{\large $q^2\to 0$}
\Text(68,27)[lb]{\large had}
\Text(0,37)[lb]{\large $A_\mu$}
\Text(145,37)[lb]{\large $A_\mu$}
\end{picture}
\end{center}

Of course we know from the beginning of this section that
\beq
\Pi_{\gamma\gamma}(q^2)\to q^2 B~~{\rm as}~q^2\to 0,
\eeq
where $B$ is some constant.
There is no reason for $B$ to be zero,  and so there is no reason 
for the derivative
of the self-energy $\Pi'_{\gamma\gamma}(0)\to B$ to be zero. Unfortunately, 
however, it is not calculable.

The incalculability of $\Pi'_{\gamma\gamma}(0)$ threatens to derail our 
precision
electroweak analysis.  However, it has been known for some time now that 
we can get at this
value by using a combination of theory tricks and experimental data.  
The first thing we do is
to rewrite $\Pi'_{\gamma\gamma}(0)$ by adding and subtracting the 
self-energy at the higher
scale $q^2=m^2_Z$:
\beq
\Pi'_{\gamma\gamma}(0)={\rm Re}\, \frac{\Pi_{\gamma\gamma}(m^2_Z)}{m_Z^2}-
\left[ \frac{ {\rm Re}\, \Pi_{\gamma\gamma}}{m^2_Z} -
\Pi'_{\gamma\gamma}(0)\right]
\eeq
The first term is calculable as computations are done at the scale 
$q^2=m^2_Z$ where
all interactions are perturbative in the SM.  The two terms in the bracket 
are not calculable, but
we will give it a name $\Delta \alpha(m_Z)$.  There are three main 
contributions to $\Delta\alpha(m_Z)$:
\beq
\Delta\alpha(m_Z)=\Delta \alpha_l(m_Z)+\Delta \alpha_{\rm top}(m_Z)
+\Delta \alpha_{\rm had}^{(5)}(m_Z)
\eeq
where
\bea
\Delta \alpha_l(m_Z) & = & 0.03150~~{\rm with~essentially~no~error} \nonumber \\
\Delta\alpha_{\rm top}(m_Z) & = & -0.0007(1)~~m_t~{\rm dependent~but~negligible} \nonumber \\
\Delta\alpha_{\rm had}^{(5)} & = & {\rm incalculable~light~hadrons~contributions} \nonumber
\eea

Fortunately, there is a way to measure $\Delta\alpha_{\rm had}^{(5)}$.  
From the optical theorem
and the methods of analytic continuation, one finds that 
\beq
\Delta\alpha_{\rm had}^{(5)}=-\frac{m_Z^2}{3\pi}\int_{4m_\pi^2}^\infty 
\frac{R_{\rm had}(q^2)dq^2}{q^2 (q^2-m^2_Z)}~~{\rm where}~
R_{\rm had}(q^2)=\frac{\sigma_{\rm had}(q^2)}{\sigma_{l^+l^-}(q^2)}.
\eeq
Therefore, to get a numerical value for $\Delta\alpha_{\rm had}^{(5)}$ 
one must integrate
over the experimental hadronic cross-section over a wide energy range.  
As soon as $q^2$ is 
significantly 
above $\Lambda_{\rm QCD}$ the theoretical cross-section can be used 
without concern.  However,
for lower $q^2$ (lower than about $5\gev$ in practice), only the 
experimental data can be used.
There are numerous experiments that contribute data for this integral 
in differing energy bins,
and it is a challenge to understand all the systematics and statistical 
errors that go into the
final number for $\Delta\alpha_{\rm had}^{(5)}$.  Many groups have gone 
through this difficult
exercise and there are many different values obtained.  The one the LEP 
Electroweak Working Group
has been using is by Burkhardt and Pietrzyk\cite{Burkhardt:2001xp}, 
who conclude that
\beq
\Delta\alpha_{\rm had}^{(5)}=0.02761 \pm 0.0036.
\eeq

We will now trade in the incalculable $\ahat$ for the calculable/measured
$\ahat(m_Z)$, which is related to the lagrangian parameters and $\Pi$'s by 
\beq
\ahat(m_Z)=\frac{\ahat}{1-\Delta \alpha(m_Z)} =\frac{e^2}{4\pi}\left[ 
1+\frac{\Pi_{\gamma\gamma}(m_Z)}{m^2_Z}\right]
\eeq
Always remember, $\ahat(m_Z)$ is an observable, which is a meaningful 
combination of many
different experiments (Thomson scattering cross-section plus integration 
over $R_{\rm had}(q^2)$),
and its experimental value is
\beq
\frac{1}{\ahat(m_Z)}=128.936 \pm 0.046.
\eeq

As for determining $v^2$ from observables, we can get it directly
and simply from the $\hat G_F$ equation
\beq
v^2=\frac{1}{\sqrt{2}\gfhat}\left[ 1 -\frac{\Pi_{WW}(0)}{m^2_W}\right].
\eeq

At this point we have $e^2$ and $v^2$ in terms of $\ahat(m_Z)$, $\mzhat$ 
and $\gfhat$, but we still
do not have the lagrangian parameter $s^2$ in terms of those three key 
observables.
To do this, we need to go to the theory prediction equation for $\mzhat$ 
and solve for $s^2$.
\beq
\mzhat^2=\frac{e^2}{4s^2c^2} v^2 +\Pi_{ZZ}(m^2_Z) \longrightarrow
s^2 c^2 = \frac{e^2 v^2}{4}\left[ \frac{1}{\mzhat^2-\Pi_{ZZ}(m_Z^2)}\right].
\eeq
After plugging in our previously obtained expressions for $e^2$ and $v^2$ 
in terms 
of observables we get after some algebra
\beq
s^2 c^2 = \frac{\pi\ahat(\mzhat^2)}{\sqrt{2}\gfhat \mzhat^2}( 1 +\delta_S)
\eeq
where
\beq
\delta_S=\frac{\Pi_{ZZ}(m_Z^2)}{m_Z^2}-\frac{\Pi_{WW}(0)}{m_W^2}
-\frac{\Pi_{\gamma\gamma}(m_Z^2)}{m_Z^2}.
\eeq
A convenient definition that I will sometimes use is
\beq
\hat s^2_0\hat c^2_0 = \frac{\pi\ahat(\mzhat^2)}{\sqrt{2}\gfhat \mzhat^2}.
\eeq
With this definition
\beq
s^2 = \hat s^2_0+\frac{\hat s^2_0 \hat c^2_0}{\hat c^2_0 -\hat s^2_0}\delta_S.
\eeq

We now have expressions for each of the lagrangian parameters in terms
of the three exceptionally well-measured 
observables $\{ \mzhat, \ahat(m_Z),\gfhat\}$
and the self-energy correction $\Pi$'s and are ready to directly compute 
the theoretical
prediction for each of the remaining observables. After some more algebra, 
which the student
should do him/herself, here are the answers:
\beq
\tp{\mwhat} =\frac{\pi\ahat(\mzhat^2)}{\sqrt{2}\gfhat \hat s^2_0}
\left[ 1-\frac{\Pi_{\gamma\gamma}(m_Z^2)}{m^2_Z}-\frac{c^2_0}{c^2_0-s^2_0}\delta_S
-\frac{\Pi_{WW}(0)}{m^2_W}+\frac{\Pi_{WW}(m_W^2)}{m^2_W}\right]
\eeq
\beq
\tp{\s2eff}= \hat s^2_0 +\frac{s^2_0 c^2_0}{c^2_0-s^2_0}
\left[ \frac{\Pi_{ZZ}(m^2_Z)}{m^2_Z}-\frac{\Pi_{WW}(0)}{m_W^2}-\frac{(c^2_0-s^2_0)}{s_0c_0}
\frac{\Pi_{\gamma Z}(m^2_Z)}{m^2_Z}-\frac{\Pi_{\gamma\gamma}(m_Z^2)}{m^2_Z}\right]
\eeq
\bea
\tp{\glhat} & = & \glhat^0 \left[ 1 -\frac{a s^2_0c^2_0}{c_0^2-s_0^2}
\frac{\Pi_{ZZ}(m_Z^2)}{m_Z^2}+\left( 1+\frac{a s^2_0c^2_0}{c_0^2-s_0^2}\right)
\frac{\Pi_{WW}(0)}{m_W^2}\right. \nonumber \\
& & \left. +a s_0 c_0 \frac{\Pi_{\gamma Z}(m^2_Z)}{m^2_Z} -\frac{\Pi_{ZZ}(0)}{m_Z^2}+a\frac{s^2_0 c^2_0}{c^2_0-s^2_0}
\frac{\Pi_{\gamma\gamma}(m_Z^2)}{m^2_Z}\right]
\eea
where 
\beq
a =\frac{-8(-1+4s^2_0)}{(-1+4s^2_0)^2+1}\simeq 0.636.
\eeq

\subsection{Summary of results}

In summary, the theoretical predictions for $\s2eff$, $\mwhat$ and $\glhat$ 
can be rewritten as
\bea
\tp{\s2eff} & = & \hat s_0^2 -(0.328)\frac{\Pi_{\gamma\gamma}(m_Z^2)}{m_Z^2}
-(0.421)\frac{\Pi_{\gamma Z}(m_Z^2)}{m_Z^2} \nonumber \\
& & -(0.328)\frac{\Pi_{WW}(0)}{m^2_W} + (0.328)\frac{\Pi_{ZZ}(m_Z^2)}{m_Z^2} 
\label{eq:s2effth}
\eea
\bea
\tp{\mwhat} & = & \mwhat^0 +(17.0\gev)\frac{\Pi_{\gamma\gamma}(m_Z^2)}{m_Z^2}
+(17.0\gev)\frac{\Pi_{WW}(0)}{m^2_W} \nonumber \\
& & +(40.0\gev) \frac{\Pi_{WW}(m^2_W)}{m^2_W}-(57.1\gev)\frac{\Pi_{ZZ}(m_Z^2)}{m_Z^2} 
\label{eq:mwhatth}
\eea
\bea
\tp{\glhat} & = & \glhat^0 + (17.5\mev)\frac{\Pi_{\gamma\gamma}(m_Z^2)}{m_Z^2}
+(22.5\mev)\frac{\Pi_{\gamma Z}(m_Z^2)}{m_Z^2} \\
& & +(101\mev)\frac{\Pi_{WW}(0)}{m^2_W}-(83.9\mev)\frac{\Pi_{ZZ}(0)}{m_Z^2} 
-(17.5\mev)\frac{\Pi_{ZZ}(m_Z^2)}{m_Z^2} \nonumber
\label{eq:glhatth}
\eea
where
\bea
\hat c^2_0\hat s^2_0 =\frac{\pi\ahat(m_Z^2)}{\sqrt{2}\gfhat \mzhat}
\eea
\bea
(\mwhat^0)^2 =\frac{\pi\ahat(m_Z^2)}{\sqrt{2}\gfhat \hat s^2_0}
\eea
\bea
\glhat^0=\frac{\ahat(m_Z^2) \mzhat}{12\hat s^2_0\hat c^2_0}
\left[ \left( -\frac{1}{2}+2\hat s^2_0\right)^2 +\frac{1}{4}\right]
\eea
with
\bea
1/\ahat(m_Z^2) & = & 128.936\pm 0.046 \\
\gfhat & = & 1.16639(1)\times 10^{-5}\gev^{-2} \\
\mzhat & = & 91.1875 \pm 0.0021\gev .
\eea

\subsection{Connection of our results to the $STU$ formalism}

A convenient parametrization of one-loop oblique corrections is given by the
$STU$ formalism\cite{Peskin:1991sw}.
In the limit that the new physics scales are much
larger than $m_Z$ one finds that the oblique corrections to all 
$Z$-pole observables are 
expressable in terms of just three universal parameters, $S$, $T$ and $U$.   
The reason why we need $m_{\rm new}\gg m_Z$ is 
because $S$, $T$ and $U$ are valid only in the approximation
that all derivatives of self-energies can be Taylor expanded to leading 
order in $m_Z/m_{\rm new}\ll 1$. (If the masses of the new particles
are close to $m_Z$, the $STU$ parameters can be augmented by additional
parameters\cite{Maksymyk:1993zm} to match the full one-loop results.)

In terms of the self-energy $\Pi$'s, the $S$, $T$ and $U$ parameters are 
\bea
S & = & \frac{\alpha}{4s^2}\left[ c^2\left( \frac{\Pi_{ZZ}(m_Z^2)}{m_Z^2}
-\frac{\Pi_{ZZ}(0)}{m_Z^2} - \frac{\Pi_{\gamma\gamma}(m_Z^2)}{m_Z^2}\right) 
\right. \nonumber \\ & & \left.  ~~~~~~
-\frac{c}{s}(c^2-s^2)\left(\frac{\Pi_{\gamma Z}(m_Z^2)}{m_Z^2}
-\frac{\Pi_{\gamma Z}(0)}{m_Z^2}\right)\right]
\eea
\bea
T = \frac{1}{\alpha}\left[ \frac{\Pi_{WW}(0)}{m_W^2}-\frac{\Pi_{ZZ}(0)}{m_Z^2}
-2\frac{s}{c}\frac{\Pi_{\gamma Z}(m_Z^2)}{m_Z^2}\right]
\eea
\bea
U & = & \frac{\alpha}{4s^2}\left[
\frac{\Pi_{WW}(m_W^2)}{m_W^2}-\frac{\Pi_{WW}(0)}{m_W^2}
-c^2\left( \frac{\Pi_{ZZ}(m_Z^2)}{m_Z^2} - \frac{\Pi_{ZZ}(0)}{m_Z^2} \right)
\right. \nonumber \\ & & ~~~~~~ \left. 
-s^2\frac{\Pi_{\gamma\gamma}(m_Z^2)}{m_Z^2}-2sc\left( \frac{\Pi_{\gamma Z}(m_Z^2)}{m_Z^2}
-\frac{\Pi_{\gamma Z}(0)}{m_Z^2}\right)\right]
\eea

The shifts in the observables $\s2eff$, $\mwhat$ and $\glhat$ can all be written as expansions in $S$, $T$ and $U$:
\bea
\Delta (\s2eff)^{\rm th} = (3.59\times 10^{-3})\, S- (2.54\times 10^{-3})\, T 
\eea
\bea
\Delta \left( \frac{\mwhat}{\mzhat}\right)^{\rm th} = -(3.15\times 10^{-3})\, S
+(4.86\times 10^{-3})\, T+(3.70\times 10^{-3})\, U
\eea
\bea
\Delta \tp{\glhat} = -(1.91\times 10^{-4})\, S + (7.83\times 10^{-4})\, T
\eea
If we plug in the expressions of $S$, $T$ and $U$ into these above equations we will find that
the result is equivalent to the expressions given by eqs.~\ref{eq:s2effth}-\ref{eq:glhatth}, with
$\Pi_{\gamma Z}(0)=0$.

\section{Cancellation of infinities}

If we think in terms of observables only, there is no issue with
infinities.  Infinities come about from intermediate steps only
when we must compute renormalized parameters in the lagrangian.
The schemes we use to bookkeep the infinities are important, especially
when one goes to higher loop order, but in the end we should remember
that they are simply not there in physical processes. Any theoretical
framework that we use must respect this obvious requirement.

We will show that the one-loop results we have computed do not introduce
infinities into observables.  The example we use for this purpose
is a top-bottom fermion loop in the vector-boson self-energies.  Before
going straight to that calculation, I wish to take a short detour
and describe Passarino-Veltman functions, which make one-loop 
analyses much more convenient.

\subsection{Passarino-Veltman functions}

In calculating vector boson self-energies we come across the same
general two-point functions over and over again.  It is useful to
define these functions carefully, make a numerical program package
to evaluate them, and never recalculate them again.  The functions
that come out from this analysis are usually called Passarino-Veltman
functions, as they were the first to systematically 
define them\cite{Passarino:1978jh}.

There are several conventions for Passarino-Veltman functions in use,
and we use the one consistent with\cite{Pierce:1996zz}:
\beq
16\pi^2 \mu^{4-n}\int \frac{d^nq}{i(2\pi)^n} \frac{1}{q^2-m^2+i\varepsilon}
=A_0(m^2)
\eeq
\beq
16\pi^2 \mu^{4-n}\int \frac{d^nq}{i(2\pi)^n}
\frac{1}{[q^2-m_1^2+i\varepsilon][(q-p)^2-m^2_2+i\varepsilon]}
=B_0(p^2,m_1^2,m_2^2)
\eeq
\beq
16\pi^2 \mu^{4-n}\int \frac{d^nq}{i(2\pi)^n}
\frac{q_\mu}{[q^2-m_1^2+i\varepsilon][(q-p)^2-m^2_2+i\varepsilon]}
=p_\mu B_1(p^2,m_1^2,m_2^2)
\eeq
\bea
16\pi^2 \mu^{4-n}\int \frac{d^nq}{i(2\pi)^n}
\frac{q_\mu q_\nu}{[q^2-m_1^2+i\varepsilon][(q-p)^2-m^2_2+i\varepsilon]} 
\nonumber
\eea
\beq
= p_\mu p_\nu B_{21}(p^2,m_1^2,m_2^2)+g_{\mu\nu}B_{22}(p^2,m_1^2,m^2_2)
\eeq

Some of these functions have poles at $n=4$, and thus have
an ``infinite'' piece proportional to 
\beq
\Delta \equiv \frac{1}{4-n}-\gamma_E+\ln 4\pi
\eeq
where $\gamma_E\simeq 0.5772$ is the Euler-Mascheroni 
constant that always accompanies
the $1/(4-n)$ pole term just as the $\ln 4\pi$ factor does.

The primitive one-point and two-point functions have analytic solutions
\beq
A_0(m^2)=m^2\left( \Delta +1 -\ln m^2/\mu^2\right)
\eeq
\bea
B_0(p^2,m_1^2,m^2_2)& = &\Delta -\int_0^1 \ln 
\frac{(1-x)m_1^2+xm^2_2-x(1-x)p^2-i\varepsilon}{\mu^2} \nonumber \\
& = & \Delta -\ln (p^2/\mu^2)-I(x_+)-I(x_-)
\eea
where
\beq
x_\pm = \frac{(p^2-m^2_2+m^2_1)\pm 
\sqrt{(p^2-m^2_2+m^2_1)^2-4p^2(m_1^2-i\varepsilon)}}{2p^2},~~{\rm and}
\eeq
\beq
I(x)=\ln (1-x)-x\ln (1-x^{-1}) -1.
\eeq
All remaining two-point functions
can be written entirely in terms of $A_0$ and $B_0$ with various 
arguments\cite{Stuart:1987tt}.
Note, $A_0(m^2)$ can also be written in terms of
$B_0$:
\beq
A_0(m^2)=m^2[1+B_0(0,m^2,m^2)]~~~{\rm since}
\eeq
\beq
B_0(0,m^2,m^2)=\Delta - \ln m^2/\mu^2,
\eeq
and so all two-point functions can be written in terms of $B_0$ functions.

One is often interested in just the infinite pieces of these functions,
as the infinite pieces dictate the renormalization group equations.
They are also helpful to check calculations,
since all infinities
must cancel in observables.   Shortly we will investigate the latter.

The relevant two-point functions in terms of 
their $\Delta$-dependent ``infinite pieces'' and their finite 
function pieces (written as lower-case) are
\beq
A_0(m^2) =  m^2\Delta +a_0(m^2) 
\eeq
\beq
B_0(p^2,m^2_1,m^2_2) = \Delta +b_0(p^2,m^2_1,m^2_2) 
\eeq
\beq
B_1(p^2,m^2_1,m^2_2) = \frac{1}{2}\Delta +b_1(p^2,m^2_1,m^2_2) 
\eeq
\beq
B_{21}(p^2,m^2_1,m^2_2) = \frac{1}{3}\Delta +b_{21}(p^2,m^2_1,m^2_2) 
\eeq
\beq
B_{22}(p^2,m^2_1,m^2_2) =
\left( \frac{m^2_1+m^2_2}{4}-\frac{p^2}{12}\right)\Delta +b_{22}(p^2,m^2_1,m^2_2)
\eeq

\subsection{Fermion loop calculation}

Now we come to our example. Let us compute the 
one-loop self energy due to a fermion loop with arbitrary vector boson 
external legs.
The basic Feynman rule notation that we use for this calculation is
\begin{center}
\begin{picture}(260,90)(-10,0)
\SetWidth{1.0}
\Photon(20,50)(70,50){3}{4}
\ArrowLine(70,50)(100,65)
\ArrowLine(100,35)(70,50)
\Text(150,45)[lb]{\large $iA\gamma^\mu (v-a\gamma_5)$}
\Text(110,60)[lb]{\large $f$}
\Text(110,30)[lb]{\large $\bar f$}
\Text(-3,40)[lb]{\large $V^\mu$}
\end{picture}
\end{center}
where $A$, $v$ and $a$ are parametrizations of the coupling.   
The fermion couplings
to a $V'$ vector boson are $A'$, $v'$, and $a'$.  
With these basic rules we
are ready to compute the one-loop function $\Pi_{VV'}(p^2)$:
\beq
i\Pi_{VV'}^{\mu\nu} = -\int \frac{d^nq}{(2\pi)^n}\, {\rm Tr}\, \left[ iA\gamma^\mu
(v-a\gamma_5)i\frac{[(\slash\!\!{q}-\slash\!\!{p})+m_2]}{(q-p)^2-m^2_2}
iA'\gamma^\nu (v'-a'\gamma_5) \frac{i(\slash\!\! {q}+m_1)}{q^2-m_1^2}\right]
\eeq
After some gamma-trace algebra and manipulations one finds that
\bea
\Pi^{\mu\nu}_{VV'} & = & \frac{AA'}{4\pi^2}\left\{ (vv'+aa')\left[ 2p^\mu p^\nu (B_{21}-B_1)
+\frac{ }{ }g^{\mu\nu} (-2B_{22}-p^2 B_{21}+p^2 B_1)\right] 
\right. \nonumber \\ & & \left. ~~~~ +\frac{ }{ } m_1m_2 (vv'-aa')g^{\mu\nu}
B_0\right\} (p^2,m^2_1,m^2_2).
\eea

As we discussed at the beginning, in precision electroweak analysis 
the only piece of the
self-energy that has a substantial influence on the observables is the 
part proportional
to $g^{\mu\nu}$: $\Pi^{\mu\nu}(p^2)=\Pi(p^2) g^{\mu\nu}+\cdots$.
As a check of our computations above, when we calculated the shifts in 
observables due
to self-energy corrections, we should check that all $\Delta$-divergences 
cancel.  After all, the 
partial width of the $Z$ boson into leptons should not divergence as 
the dimensions approach
$n\to 4$ in the calculation.  

For fermion self-energies, we can check for finiteness of the theory 
predictions given the
expressions above.  The $\Delta$-divergence part of $\Pi_{VV'}$ is
\beq
\Pi^\Delta_{VV'}(p^2)=\frac{AA'}{4\pi^2}\left\{ (vv'+aa')\left( -\frac{1}{2}(m^2_1+m^2_2)+\frac{p^2}{3}\right)
+(vv'-aa')m_1m_2\right\}\Delta
\eeq
For the top-bottom quark doublet, we can compute these $\Delta$-divergence 
pieces.  The nonzero
contributions are
\beq
\Pi^\Delta_{ZZ}(m_Z^2) = m_Z^2  \sum_{i=t,b} \frac{e^2}{4s^2c^2}\left[ (T^3_i-2Q_i s^2)^2+(T^3_i)^2\right]\, \Delta 
\eeq
\beq
\Pi^\Delta_{\gamma\gamma}(m_Z^2)=m_Z^2\sum_{i=t,b} (eQ_i)^2\, \Delta 
\eeq
\beq
\Pi^\Delta_{\gamma Z}(m_Z^2)=m_Z^2 \sum_{i=t,b} \frac{e^2 Q_i}{2sc}(T^3_i-2Q_i s^2)\,  \Delta
\eeq
\beq
\Pi^\Delta_{WW}(m_W^2) = m_W^2  \frac{e^2}{4s^2}\, \Delta
\eeq
Substituting these expressions into the eqs.~\ref{eq:s2effth}-\ref{eq:glhatth}, 
one finds that
all $\Delta$-divergent terms cancel identically, as they should.

\section{Note on the utility of these techniques}

Finally, I would like to emphasize how useful the oblique correction 
calculations can
be to research.  Suppose you have a beyond-the-SM (BSM) theory that for 
one reason
or another induces small corrrections to the theoretical predictions
of observables compared to the SM, and all those corrections can be 
expressed entirely
in terms of vector boson self-energies.  In this case, one can perform a 
rigorous precision
electroweak analysis of the theory by following the techniques
described above.

This case is applicable when the set of beyond-the-SM states under
consideration cannot couple directly to the final state fermions.
An example of this would be split supersymmetry, where the gauginos
and higgsinos are light but the sfermions are 
decoupled\cite{splitsusy,Wells:2003tf}.  
In that case,
there is no way to couple the higgsinos and gauginos directly to 
the final states, because there are no superpartners of the fermions
to complete the vertex.  Thus, oblique corrections as discussed here
are the way to go\cite{Martin:2004id}.  

Another example is in some strongly coupled theories that have 
pseudo-Nambu-Goldstone bosons with gauge quantum numbers but no
flavor quantum number to complete a sizeable vertex correction.
Also, if we have a collection of numerous beyond-the-SM states, almost all
will couple to the vector bosons in some way (representations under
their gauge groups), whereas only a few at most will couple to any
given final state. Thus, we expect oblique corrrections to often
be the most important corrections even when the set of beyond-the-SM
states do contain fields that couple to external fermions.

Once we decide that oblique corrections are the appropriate set of
corrections to apply to our observables in a beyond-the-SM setting, 
we should follow these practical steps to perform a precision
electroweak analysis.

First, one must get control of the SM observable predictions. 
The full SM precision electroweak analysis,
with all vertex corrections, and higher-order QCD corrections, etc., 
is a very involved process.
If your goal is not to redo this analysis, you can find an analysis 
that you trust.  For example,
the values of the observables given by the LEP Electroweak Working 
Group~\cite{LEPEWWG}
for reference values
of the input parameters ($\alpha_{\rm had}$, $m_t$, $\alpha_s$, $m_h$, etc.) 
can 
often be used for the starting point. Likewise, the many precision electroweak
programs available on the market, such as ZFITTER\cite{Bardin:1999yd}, 
can be used.
Using $a_i$ as a generic symbol for a 
SM input parameter, one
finds
\beq
{\cal O}^{\rm SM}(\{ a_i\}) = {\cal O}( \{ a^{\rm ref}_i\})+
\sum_i c_i (a_i - a^{\rm ref}_i)+\cdots
\eeq
The coefficientis $a_i$ can be found in many 
publications\cite{Ferroglia:2002rg}.
The key to this step is to compute or get
the state-of-the art computation for
${\cal O}^{\rm SM}(\{ a_i\}) $ given various $a_i$ inputs.

Second, compute the self-energy corrections due to the new BSM states.  
Using $\eta_j$
as a generic symbol for a BSM input parameter, we can write our full 
expression for
the observable ${\cal O}$ in terms of both the SM and BSM input parameters
\beq
{\cal O}^{\rm th}(\{ a_i\},\{ \eta_j\}) = {\cal O}^{\rm SM}(\{ a_i\}) + 
\delta{\cal O}^{\rm BSM}(\{ a_i\},\{\eta_j\})
\eeq
Notice, our notation illustrates that we are viewing the BSM contributions 
as small shifts
to the SM predictions.  This is likely to be true given the apparently good 
agreement the SM has with the
precision electroweak data.  If this is not true, a full 
renormalization procedure {\it de novo} must
be carried out to have confidence in the result.

Third, set up a  $\chi^2$ analysis of the full BSM theory:
\beq
\chi^2 = \sum_k \frac{({\cal O}^{\rm th}_k(\{ a_i\},\{ \eta_j\})-{\cal O}_k^{\rm expt})^2}{(\Delta {\cal O}_k)^2}
\eeq
Minimizing the $\chi^2$ enables one to find the best-fit values of the 
$\{ a_i\}$ and $\{ \eta_j\}$ parameters.  If the $\chi^2$ per 
degree of freedom is good,
the theory survives.  An interesting application of this type of 
analysis is to add additional
new physics contributions (turning on some $\eta_j$ contributions) 
and see if the SM Higgs mass (one of the $a_i$ parameters) can be 
significantly heavier at the
95\% confidence level than
its value of about $200\gev$ in the pure SM analysis 
(all $\eta_j$ decoupled)\cite{Peskin:2001rw}.

The student should look out for two common exceptions: nonuniversal
corrections to the $Zb\bar b$ vertex and $Z'$ corrections.  $Zb\bar b$
vertex corrections are often present in beyond-the-SM theories that
treat the 3rd family special in any way, and one must be careful
to take them into account.  Luckily, these corrections are often
finite, gauge invariant corrections all by themselves and can be 
inserted into the analysis rather easily.  As for $Z'$, many of its
effects are more
akin to our beginning tree-level analysis than one-loop corrections.
Using the techniques of these lectures, combined with an understanding of
all ways a $Z'$ boson can interact with SM states, enables one to do an
analysis of $Z'$ implications to precision electroweak observables\cite{Babu:1997st}.  
I recommend the student make up
a $Z'$ boson with his/her favorite couplings and compute all the corrections
to observables as an exercise.  

Finally, at the end of the lectures we spent some time with the
full numerical results of the precision electroweak fits to the
standard model.  Many of the figures that I showed were from
the annual LEP Electroweak Working Group report\cite{LEPEWWG}.
I encourage students who are interested in delving deeper
into the technical aspects of this topic to read carefully this important
document.

\section*{Acknowledgements}
I wish to thank John Terning, Carlos Wagner and Dieter Zeppenfeld, the
organizers of TASI 2004, for the invitation to lecture at this stimulating
summer school.  I thank K.T. Mahanthappa, the staff, and other locals
at University of Colorado for their support and hospitality. And finally,
I thank the students and other lecturers 
of TASI 2004 for contributing to such an energetic atmosphere.

\end{document}